\documentclass[a4paper]{article}
\usepackage{color,graphicx,bm,multirow,authblk}
\usepackage[margin=2.7cm]{geometry}
\usepackage[UKenglish]{babel}
\usepackage[sort&compress,numbers]{natbib}
\usepackage{url}
\usepackage{verbatim}

\usepackage{setspace}
\title{A Review of Methods for the Analysis of the Expected Value of Information}
\author[1]{Anna Heath}
\author[1]{Ioanna Manolopoulou} 
\author[1]{Gianluca Baio}
\affil[1]{Department of Statistical Science, University College London}

\begin{document}
\maketitle
\abstract
Over recent years Value of Information analysis has become more widespread in health-economic evaluations, specifically as a tool to perform Probabilistic Sensitivity Analysis. This is largely due to methodological advancements allowing for the fast computation of a typical summary known as the Expected Value of Partial Perfect Information (EVPPI). A recent review discussed some estimations method for calculating the EVPPI but as the research has been active over the intervening years this review does not discuss some key estimation methods. Therefore, this paper presents a comprehensive review of these new methods. We begin by providing the technical details of these computation methods. We then present a case study in order to compare the estimation performance of these new methods. We conclude that the most recent development based on non-parametric regression offers the best method for calculating the EVPPI efficiently. This means that the EVPPI can now be used practically in health economic evaluations, especially as all the methods are developed in parallel with \verb"R" functions to aid practitioners.

\footnote{Financial support for this study was provided in part by grants from EPSRC [Anna Heath] and Mapi [Dr. Gianluca Baio]. The funding agreement ensured the authors' independence in designing the study, interpreting the data, writing, and publishing the report.}

%%%3,500 words -MAX - MDM or 6000 words - MAX - PharmacoEconomics
\section{Introduction}
Health economic models are used to predict the costs and health effects of competing healthcare technology decision options. Model input parameters are informed by real-world study evidence and are therefore rarely known with certainty. The typical method for assessing input uncertainty is Probabilistic Sensitivity Analysis (PSA), i.e.\ Monte Carlo or bootstrap sampling from the input parameter probability distribution coupled with model evaluation in order to generate samples of predicted costs and effects. PSA is a fundamental component of any health economic evaluation and is recommended by Health Technology Assessment (HTA) bodies such as the National Institute for Health and Care Excellence (NICE) in the UK \cite{Claxtonetal:2005}, across other countries in Europe \cite{eunethta:2014} and beyond \cite{CADTH:2006,Australia:2008}.

A fully decision-theoretic approach to PSA %, which avoids the shortcomings of CEACs,
is based on the analysis of the \textit{value of information} (VoI) \cite{Howard:1966}, an increasingly popular method in health economic evaluation \cite{FelliHazen:1998,FelliHazen:1999,Claxton:1999,Claxtonetal:2001,Adesetal:2004,BrennanKharroubi:2005,Briggsetal:2006,Fenwicketal:2006,Lotteetal:2013, Keisleretal:2014}.
%In this approach, the \textit{ideal} scenario, which would occur if we were able to resolve the uncertainty in the parameters, e.g.\ by observing a very large amount of data, is compared with the \textit{actual} decision-making process, in which many of the parameters are known with limited precision and thus uncertainty is averaged~out. In other words, the VoI 
A VoI analysis quantifies the expected value of obtaining \textit{perfect} information about the underlying model parameters. Because it is usually impossible to obtain perfect information, this analysis places an upper bound on the cost of additional research aimed at reducing this uncertainty. %If the additional research is projected to have a cost greater than the expected value of perfect information, then the decision-maker can base their decision on the current evidence, which is deemed to be sufficient. 

The main advantage of the analysis of the VoI is that it directly addresses the potential implications of current uncertainty, not only in terms of the likelihood of modifying the current decision in light of new and more definitive evidence, but also in terms of the opportunity cost of the incorrect decision. If this cost is low, there is little value in agonising about the decision, even for a low probability of cost-effectiveness, as the implicit penalty is negligible if the decision turns to be wrong in the face of new evidence. Therefore, the probability of cost-effectiveness %(the CEAC) 
alone can dramatically overstate decision sensitivity. For this reason, it has been advocated that VoI measures should also be presented when representing decision uncertainty \cite{FelliHazen:1998,Baio:2012,Meltzer:2001,Campbelletal:2014}.

Despite the useful features of a VoI analysis, its uptake in health economic evaluation has been slow. VoI analysis has been hindered by two factors in the main. First, the interpretation of the expected value of information is %less straightforward than that of the CEAC, which shows the probability of cost-effectiveness as a function of the willingness to pay threshold and, as such, lies on the interval [0,1]. Conversely, VoI 
not straightforward as VoI measures are theoretically unbounded, making interpretation more challenging. 

Secondly, and perhaps more importantly, VoI measurements can be computationally costly unless restrictive assumptions are made \cite{Wilson:2015}. 
Practically, it is typically easy to calculate numerically (but not analytically) the expected value of learning \textit{all} the model parameters perfectly. This is known as the (overall) Expected Value of Perfect Information (EVPI).  %, particularly under a Bayesian approach based on simulations (e.g.\ MCMC algorithms). 
However, this has often little practical use, as it will be rarely possible to learn \textit{all} the model parameters at once in a new study. Thus, the decision-maker is usually interested in the expected value of information about subsets of parameters. This will indicate which parameters are driving decision uncertainty, and thus where future research may add value. This subset analysis is called the \textit{Expected Value of Perfect Partial Information} (EVPPI), and it is this analysis that is computationally costly. %While theoretically this can be seen as special case of the EVPI, the actual calculation is usually analytically intractable and the simulation based approach is computationally very challenging. 
Therefore, significant effort has been invested in developing more efficient methods for computing EVPPI \cite{FelliHazen:1998,StrongOakley:2013,Brennanetal:2007,Sadatsafavietal:2013,Oakley:2002,
StrongOakley:2014,Madanetal:2014}. 

A relatively recent review \cite{CoyleOakley:2008} discussed some of these methods. However, research in this area has been active over the last few years. We present and evaluate these recent methodological advances, concluding that the most recent developments made in \cite{StrongOakley:2014} offer the best method for approximating the EVPPI. 

We begin by formalising the EVPPI calculation and briefly discussing the methods from the previous review. We then %In \S \ref{single}, we 
present two methods for single-parameter EVPPI before discussing the most recent developments based on non-parametric regression. % \S \ref{non-parametric}. 
Then, %In \S \ref{results}, 
we compare the performance of these methods using a fictional but realistic example of economic modelling. From our analysis, it appears that all methods produce comparable results and have similar standard errors and computational time. Finally, we present some discussion to suggest that, although they are still affected by some computational limitations, non-parametric regression methods offer the most suitable EVPPI approximation method.

%%%%several important developments have been made since. This review will begin by briefly discussing the method presented in this previous review. It will then present the methodological advances made since this previous review paper. For each of these new approximation methods the algorithm is presented in full. Next, in \S \ref{results}, the performance of these new approximations is tested by comparing the approximate EVPPI value with an older method. \emph{We conclude that, for our case study, all the new methods adequately approximate the underlying EVPPI value and can therefore be used to approximate the EVPPI. }

%%%%We show that the standard error for all the methods is approximately equivalent and therefore the main differences between the methods are therefore the ease of implementation and the speed of computation. For all methodological advancements, R code is provided by the original authors and thus these algorithms can all be used by practitioners without an in-depth understanding of the underlying theory. Therefore, all methods can be implemented easily. In \S\ref{discussion}, we note the dangers associated with naively applying some of the methods and conclude that due to this the most recent developments in \cite{StrongOakley:2014} offer the current best method for approximating the EVPPI. Finally, we suggest some avenues of further research into EVPPI approximation methods. 

\section{Notation and basic concepts}
Health economic decision making aims at determining the optimal intervention in terms of costs and health effects, among a number $T$ of options, often set to $T=2$. In general we can consider the case where a larger number of interventions is available, although in practice it is uncommon that $T > 5$. Usually, we consider $t=0$ as the standard intervention currently available for treatment of a specific condition and $t=1,\ldots,T$ as a (set of) new option(s) being assessed.

Each alternative $t=0,\ldots,T$ is valued by means of a \textit{utility function}, typically defined in terms of the monetary net benefit
\[ nb_t = ke_t - c_t. \]
Here, $(e_t,c_t)$ is the health economic response, typically subject to individual variability expressed by a joint probability distribution $p(e,c\mid\bm\theta)$, indexed by a set of relevant parameters $\bm\theta$, 
%In addition, $k$ is the \textit{willingness to pay}, which is used to put the cost and effectiveness measures on the same scale, i.e. $k$ is the amount of money that the decision maker is willing to pay to increment the benefits by one unit. 
while $k$ is the \textit{willingness to pay}.

In a full Bayesian setting, a complete ranking of the alternatives is obtained by computing the overall expectation of this utility function over both individual variability and parameter uncertainty
\[ \mathcal{NB}_t = k\mbox{E}[e_t] - \mbox{E}[c_t], \]
i.e.\ the expectation here is taken with respect to the joint distribution $p(e,c,\bm\theta) = p(e,c\mid\bm\theta)p(\bm\theta)$. The option $t$ associated with the maximum overall expected utility $\mathcal{NB}^*=\max_t \mathcal{NB}_t$ is deemed to be the most ``cost-effective'', given current evidence. 

However, when performing PSA, we consider the expected utility taken with respect to individual variability only (i.e.\ as a function of the parameters $\bm\theta$)
\begin{eqnarray*}
\mbox{NB}_t(\bm\theta) = k\mbox{E}[e_t\mid \bm\theta]-\mbox{E}[c_t\mid\bm\theta],
\end{eqnarray*} 
which, in line with \cite{Baio:2012}, we term the ``known-distribution'' net benefit. In this case, the expectation is taken with respect to the conditional distribution $p(e,c\mid \bm\theta)$. Thus, while $\mathcal{NB}_t$ is a deterministic quantity, $\mbox{NB}_t(\bm\theta)$ is a random variable (randomness being induced by uncertainty in the model parameters $\bm\theta)$ and %the following relationship holds 
$\mbox{E}\left[\mbox{NB}_t(\bm\theta)\right]  = \mathcal{NB}_t.$

The vector of parameters can in general be split in two components $\bm\theta=(\bm\phi,\bm\psi)$, where $\bm\phi$ is the sub-vector of parameters of interest (i.e.\ those that could be investigated further) and $\bm\psi$ are the remaining ``nuisance'' parameters. To calculate the EVPPI, we compute a weighted average of the net benefit for the optimal decision at every point in the support of $\bm\phi$ after having marginalised out the uncertainty due to $\bm\psi$:%. The mathematical definition is given as 
\begin{equation}\mbox{EVPPI} = \mbox{E}_{\bm\phi}\left[\max_t \mbox{E}_{\bm\psi\mid\bm\phi} \left[\mbox{NB}_t(\bm\theta)\right]\right] - \max_t \mbox{E}_{\bm\theta} \left[\mbox{NB}_t(\bm\theta)\right], \label{EVPPI}
\end{equation} 
where $\max_t \mbox{E}_{\bm\psi\mid\bm\phi}\left[\mbox{NB}_t(\bm\theta)\right]$ is the value of learning $\bm\phi$ with no uncertainty. Of course, this is only an idealised state of nature, as we will hardly even be in the position of completely eliminating the uncertainty on $\bm\phi$; thus, we then average over its current probability distribution.

\section{A review of available methods for the computation of the EVPPI in health economic evaluations}
It is rarely possible to calculate the EVPPI analytically as the known distribution net benefit is often a complicated function of the underlying parameters. Additionally, it is generally challenging to calculate analytically the expectation of a maximum, as is required in the first term of (\ref{EVPPI}). For this reason, a wealth of papers have been dedicated to finding accurate approximation methods for the~EVPPI.

\subsection{EVPPI approximation methods requiring additional sampling or underlying assumptions} 
The most recent methodological review of these methods \cite{CoyleOakley:2008} presented four approximation procedures for the EVPPI. Two of these methods are computationally intensive and derive directly from the construction of the first term in (\ref{EVPPI}). The other two methods can only be used when the known-distribution net benefit function fulfils certain conditions, generally involving some form of linearity in the parameter sets and/or simplifying distributional assumptions. These methods are considerably faster in terms of computational time.

The most important computationally intensive method is a two-step Monte Carlo procedure that was first formalised in \cite{Brennanetal:2007}. For this method, Monte Carlo sampling is used to approximate both the inner and outer expectations in the first term of (\ref{EVPPI}). The conditional expectation $\mbox{E}_{\bm\psi\mid\bm\phi}[\mbox{NB}_t(\bm\theta)]$  must be calculated using a nested simulation as it is dependent on $\bm\phi$. Therefore, the main limitation of this method is the immense computation power needed to calculate an accurate estimate for the EVPPI. Although theoretically Monte Carlo simulation can give any desired accuracy, the number of nested simulations necessary to produce accurate estimates would generally require an infeasibly long computation time, making this method impractical, despite theoretically offering good estimation properties. The second expensive algorithm uses numerical integration to reduce the number of nested samples needed. However, it does not offer a significant computational saving when compared to the two-step Monte Carlo method and is impractical for high-dimensional subsets of $\bm\phi$.

The faster, restrictive methods for EVPPI calculation include the \textit{unit loss integral} approximation, which assumes the parameters of interest $\bm\phi$ are approximately normal and uses this approximation to derive analytical results. The second restrictive approximation was first developed in \cite{FelliHazen:1998} under the assumptions that the net benefit is a linear function in the ``nuisance'' parameters $\bm\psi$ and that these are independent of the relevant parameters $\bm\phi$ (although these conditions can be slightly relaxed \cite{Brennanetal:2007,Madanetal:2014}). These assumptions ensure that it is possible to calculate the EVPPI using the samples from the parameters distribution, which are already available from PSA. However, this approximation deteriorates if the independence assumption is not valid.

More recently, Madan \textit{et al.}\ \cite{Madanetal:2014} explored some re-parametrisations and approximations of the net benefit such that this method can be used when its original formulation does not conform to these restrictive conditions. For example, if the net benefit is non-linear in $\bm\psi$ but $\bm\psi$ and $\bm\phi$ are independent, it would then simply be possible to re-parametrise the problem so that $\mbox{NB}_t$ is linear in a new parameter set $\bm\psi'=f(\bm\psi)$. 

All of these methods rely on additional information about the model --- either to perform more sampling or to check whether it conforms to certain conditions --- making EVPPI calculation challenging, both methodologically and computationally. However, since the review of \cite{CoyleOakley:2008}, significant developments have been made allowing the EVPPI to be calculated quickly using the data already available as part of a standard PSA procedure with no knowledge of the underlying model structure.

\subsection{Approximations for Single Parameter EVPPI} \label{single}
Recently, two methods have been specifically developed to approximate the EVPPI for a single parameter, $\phi$, which rely on the same underlying ideas: i) provided the optimal decision remains constant, the first term for the EVPPI can be approximated by the average known-distribution net benefit value within this set; and ii) if a treatment option is optimal at one point in the parameter set, it is still optimal for parameter values ``close to'' that original point. For a set of $S$ simulated parameter values $(\bm\theta^1, \dots, \bm\theta^S)$, it is possible, therefore, to approximate the first term in (\ref{EVPPI}) provided there is some assurance that the optimal decision remains constant within the sets of interest.
 
\subsubsection{Strong and Oakley} 
The first method \cite{StrongOakley:2013} is based on the idea that if the parameter space is split into ``small'' subsets, the optimal decision is unlikely to change within each of these subsets, as it is reasonable to assume it is locally constant. 

Thus, to calculate the EVPPI, it is necessary to determine subsets of the simulated values of $\mbox{NB}_t(\bm\theta)$ for which the simulated values of $\phi$ are similar. Practically, this is achieved by reordering the observed known-distribution net benefit values so that they have to same order as the simulated values for $\phi$ --- note that since it is assumed that $\phi$ is a scalar, ordering is trivial. This list of ordered values is then split into $M$ small sub-lists of length $L=\frac{S}{M}$. Within each sub-list the average known-distribution net benefit is calculated for each treatment option and the maximum within each subset is used as an estimate for the first term in equation (\ref{EVPPI}), in correspondence of the simulated value for $\phi$. 

Therefore, %to calculate the single-parameter EVPPI using the Strong and Oakley method, 
the following strategy can be used:
\begin{enumerate}
\item Sample $S$ values from the distribution of the parameters $\bm\theta$; we indicate these %values 
as $\bm\theta^{1},\ldots,\bm\theta^{S}$.
\item At each iteration $s=1,\ldots,S$, use the simulated parameter vector $\bm\theta^{s}$ to compute the estimated known-distribution net benefit $\mbox{NB}_t(\bm\theta^{s})$ for each treatment option $t$.
\item Sort the simulated values of the parameter of interest in ascending order --- for simplicity, we write the sorted vector as $\varphi^1,\ldots,\varphi^S$, where $\min\phi=\varphi^1 <\ldots<\varphi^S=\max\phi$. 
\item Re-order the estimated known-distribution net benefits as $\mbox{NB}_t(\varphi^1),\ldots,\mbox{NB}_t(\varphi^S)$, where $\mbox{NB}_t(\varphi^s)$ is the net benefit corresponding to the $s-$th ordered simulated value of $\phi$.
\item Split the ordered list of known-distribution net benefits into $m=1,\ldots,M$ sub-lists $\mathcal{L}_m$ and compute the average in each of the sub-lists and for each treatment $t$ to obtain
\[ \widehat{\mbox{NB}_t^m}(\bm\theta) = \frac{1}{L} \sum_{j \in \mathcal{L}_m} \mbox{NB}_t(\varphi^j).\]
\item Compute $\displaystyle\widehat{\mbox{EVPPI}} = \frac{1}{M}\sum_{m=1}^M\max_t \widehat{\mbox{NB}_t^m}(\bm\theta).$
\end{enumerate}

The approximation given by this method is highly sensitive to the value of $M$. If $M$ is large, the EVPPI is close to the overall EVPI estimate. At the other end of the scale, if $M$ is small, the EVPPI is approximately 0. Strong and Oakley \cite{StrongOakley:2013} use a normal approximation to estimate the upward bias of the EVPPI calculated using this method. They suggest choosing $M$ to be the largest number of subsets such that the upward bias falls below a pre-specified threshold value, say 0.1. In this manner, the upward bias of the EVPPI estimate is controlled but, hopefully, the downward bias, present for small values of $M$, is avoided. 

\subsubsection{Sadatsafavi et al.}
This second method for single parameter EVPPI \cite{Sadatsafavietal:2013} can be thought of as an extension to the previous one, although both were developed concurrently. In this case, the ordered list 
\[ \mbox{\textbf{NB}}(\bm\theta) = \left(\begin{array}{ccc}
\mbox{NB}_0(\varphi^1) & \cdots & \mbox{NB}_0(\varphi^S) \\
\mbox{NB}_1(\varphi^1) & \cdots & \mbox{NB}_1(\varphi^S) \\
\cdot & \cdot & \cdot \\
\mbox{NB}_T(\varphi^1) & \cdots & \mbox{NB}_T(\varphi^S) \\
\end{array}\right)
\]
is split in an informed manner. The algorithm searches for the point(s) at which the parameter of interest is directly responsible for a change in the optimal decision. Although, in general, variations in a parameter may induce a change in the optimal decision a large number $M$ of times, in practice a single parameter is unlikely to modify it more than once or twice over the range of values selected for the willingness-to-pay. Provided we search for the correct number of decision changes for the parameter in question, the method calculates the true EVPPI values, asymptotically as the number of PSA samples goes to infinity. 

The algorithm developed by Sadatsafavi \textit{et al.} is identical to that of Strong and Oakley described in the previous section in steps 1-4, but then proceeds as follows:
\begin{enumerate}
\setcounter{enumi}{4}

\item Split the ordered list of net benefits into $m=1,\ldots,M$ sub-lists $\mathcal{L}_m$. In this case, the list of known-distribution net benefit values is split at $M$ points, $\left(\varphi^{\mathcal{L}_1},\ldots,\varphi^{\mathcal{L}_M}\right)$, known as the \textit{segmentation vector}. This means that the $m-$th sub-list contains all the net benefit values calculated with the value $\varphi$ such that $\varphi^{\mathcal{L}_m} \leq \varphi< \varphi^{\mathcal{L}_{m+1}}$.
\item Calculate and store $\displaystyle\widehat{\mbox{NB}_t^m}(\bm\theta) = \frac{1}{L} \sum_{j \in \mathcal{L}_m} \mbox{NB}_t(\varphi^j)$.
\item Maximise over all possible segmentation vectors to compute 
\begin{eqnarray*}
\widehat{\mbox{EVPPI}} = \max_{(\varphi^{\mathcal{L}_1},\ldots,\varphi^{\mathcal{L}_M})}\frac{1}{M}\sum_{m=1}^M\max_t \widehat{\mbox{NB}_t^m}(\bm\theta).
\end{eqnarray*}
\end{enumerate}

It is important to note that this method is also sensitive to the choice of $M$. It can be shown that the EVPPI estimate is asymptotically unbiased if we search for the true underlying number of decision changes. Therefore, Sadatsafavi \textit{et al.} suggest a systematic method for choosing $M$, based on a visual tool for determining the number of decision changes. This tool plots the cumulative sum of the differences between the known-distribution net benefits for the two treatment options. When the optimal decision changes, so does the sign of the difference between the two treatment options. Thus, a change in optimal decision is shown by an extremum for this cumulative sum. An example of this visual tool is given in the supplementary material. Despite the usefulness of the visual tool, it can be difficult to identify extrema making EVPPI estimates using this method unreliable in certain situations.

Both these single-parameter estimation methods are dependent on their inputs making ``black-box'' calculations difficult. In addition, since the publication of these two methods, further research has been developed that allows to calculate multi-parameter EVPPI using the PSA samples and non-parametric regression methods.

\subsection{Non-parametric regression for EVPPI calculations}\label{non-parametric}
%\note[AH]{This entire section has changed and the original is commented out in the \LaTeX code. This is for two reasons, firstly I've realised my first explanation was wrong, secondly I think it's pointless to talk about Meta-Models. Better to leave it in terms on regression??}

%%%\add[AH]{The first use of non-parametric regression for EVPPI estimated the health-economic model itself.} A health-economic model calculates the net benefit value for each treatment option as a function of the parameter values. In complicated models, such as a Markov model, this calculation can be complicated and take a non-trivial computation time. Therefore, for these complex health-economic models, the computation of the EVPPI via the 2 step Monte Carlo method is often infeasible. \remove[AH]{Firstly, the parameters must be sampled, via bootstrapping or MCMC, and then furthermore the health-economic model must be run for each different parameter subset. }Oakley \cite{Oakley:2002} suggests \change[AH]{using non-parametric regression to approximate the health-economic model,}{therefore,} creating an approximate health-economic model that estimates the net benefit with little or no computational effort.
The basic idea proposed initially by Oakley \citep{Oakley:2002} and developed further by Strong et al.\ \cite{StrongOakley:2014} is that regression methods can be used to approximate the EVPPI. Specifically, Strong et al.\ \cite{StrongOakley:2014} demonstrate that the known-distribution net benefit can be approximated, based on sampled parameter values. In particular, since health economic models are likely to be complicated functions of the underlying parameters, flexible regression methods should be used, in order to limit the number of assumptions on the relationship between the known-distribution net benefit and the parameters.

In a nutshell, for each simulated value of the parameters $s$, Strong et al. propose to approximate the inner conditional expectation in (\ref{EVPPI}) as
\begin{eqnarray*}
\mbox{NB}_t(\bm\theta^s) = \mbox{E}_{\bm\psi\mid\bm\phi^s}\left[\mbox{NB}_t(\bm\phi^s,\bm\psi^s)\right]+\varepsilon^s,
\end{eqnarray*}
with $\varepsilon^s \sim \mbox{Normal}(0,\sigma_\varepsilon^2)$.  Furthermore, the conditional expectation can be thought of as a function of $\bm\phi$ only, as the \emph{conditional} expectation is dependent on the value of $\bm\phi$. Therefore, the problem can be formulated as 
\begin{eqnarray*}
\mbox{NB}_t(\bm\theta^s) = g_t(\bm\phi^s)+\varepsilon^s,
\end{eqnarray*} 
for which the $\bm\phi^s$ are ``observed'' values (obtained via Monte Carlo sampling) for the independent variables and $\mbox{NB}_t(\bm\theta^s)$ is the ``observed'' dependent variable. 

Strong et al.\ \cite{StrongOakley:2014} provide \verb"R" code \cite{Strong:2012:Code} for two alternative non-parametric regression methods: Generalised Additive Models (GAMs) \cite{Hastie:1990} and Gaussian Processes \cite{Rasmussen:2006}. Both these methods offer a large amount of flexibility but other regression methods could be used. In a general sense, the EVPPI estimate is calculated by:
\begin{enumerate}
\item Sample $S$ values $\left(\bm\theta^{1},\ldots,\bm\theta^{S}\right)$ from the distribution of the parameters.
\item Calculate the net benefit $\mbox{NB}_t(\bm\theta^{s})$ for each observed parameter vector $\bm\theta^{s}$ and treatment option~$t$.
\item For each treatment option $t$, fit a regression curve with $\bm\phi^{s}$ as the observed ``covariates'' and $\mbox{NB}_t(\bm\theta^s)$ as the observed ``dependent'' variable, for the $s-$th ``unit'' of analysis (i.e.\ simulations).
\item For each treatment $t$, find the $s-$th fitted values $g_t(\bm\phi^{s})$ by inputting the observed values $\bm\phi^{s}$ into the regression curves.
\item Calculate $\displaystyle\widehat{\mbox{EVPPI}} = \frac{1}{S}\sum_{s=1}^S \max_t g_t(\bm\phi^{s}) - \max_t \frac{1}{S} \sum_{s=1}^S g_t(\bm\phi^{s})$
\end{enumerate}
This non-parametric regression method was published most recently and thus represents the current state of the research into estimation methods for the EVPPI. However, in spite of the relative complexity of this method, ``black-box'' calculations are possible using automatically chosen hyper-parameters for both non-parametric methods as they require little additional inputs from the user. This allows for the creation of general purpose functions \cite{Strong:2012:Code} and the web application \textit{Sheffield Accelerated Value of Information} (\texttt{SAVI} \cite{SAVI:2014}) which allows users to calculate EVPPI. This web app also produces some standardised output, graphics, tables, that can be easily presented in formal reports. All these methods are also implemented in the \texttt{R} package \texttt{BCEA} \citep{BCEA:2013,BaioBCEA:2012}.

\section{Comparison of methods} \label{results}
This section compares the EVPPI estimates given by the methods discussed in this review and the two-step Monte Carlo procedure. Firstly, a case study pertaining to influenza infection vaccination is presented. We then discuss how the EVPPI estimates were calculated with regards to the input values for the single-parameter methods. Next, we present the results of our analysis; all five methods give approximately the same EVPPI estimate, with the same standard error for the four new methods. Therefore, the methods should be compared on their ease and speed of implementation and applicability.
\subsection{Case Study - Vaccine}
A Bayesian health-economic model is proposed to analyse the effect of the influenza vaccine on health-outcomes and costs. This example is relatively simple but complex enough to be intractable analytically. Therefore, the parameters values must be sampled from their joint posterior distribution using Markov Chain Monte Carlo (MCMC) methods. 

Two treatment options are considered, either the vaccine is available to the population ($t=1$), or not ($t=0$). If a patient gets influenza, they are treated with anti-viral drugs and will often visit the GP. Complications may occur, including pneumonia and hospitalisation -- in which case there will be some indirect costs such as time off work. The cost of the treatment is the acquisition cost of the drugs, the time in hospital, the GP visits and the cost of the vaccine. The benefit of the treatment is measured in QALYs (Quality Adjusted Life Years), where each adverse effect contributes negatively to the benefit.

There are 28 key parameters in the model representing the probability of infection, the reduction in risk due to the vaccine, the occurrence of complications, the monetary costs of the interventions and the QALY loss due to different health states. A detailed discussion of the full model parametrisation is presented in \cite{Baio:2012}. The single parameter EVPPI was calculated for all of these 28 parameters using all the methods outlined.
\subsection{Analysis}
This analysis compares the methods presented by estimating single parameter EVPPI, as some of the methods are only suitable for single-parameter estimation. As previously discussed, both single parameter estimation methods have input values that greatly affect the estimation. For the Strong and Oakley method, the number of subsets must be specified by the user and for the Sadatsafavi \textit{et al.}\ method, the number of decision changes for each parameter is needed. The estimates are \emph{very} sensitive to this choice. In fact, for the Sadatsafavi \textit{et al.} method, if the visual tool indicates that the parameter does not affect the decision, this is equivalent to indicating that the EVPPI is equal to 0. This means that in some cases visual indications are used to determine the EVPPI rather than objective analysis. 

To determine these influential input values, the suggested visual tool was used for each of the 28 parameters in the Vaccine case study\footnote{The supplementary material gives a more in depth explanation of this visual tool and implements it for some of the parameters}. It was decided that 23 out of the 28 parameters change the optimal decision once, the other parameters have no effect -- these correspond to an EVPPI equal to 0 in Table \ref{EVPPI-table}. For the Strong and Oakley \cite{StrongOakley:2013} method, the upward bias of the EVPPI estimate was assessed for all the parameters in the data set. The number of subsets was taken as the largest number of subsets such that the upward bias remains below 0.1. The magnitude of the bias differed for each parameter, and therefore the number of subsets needed to reduce the upward bias below 0.1 was between 20 and 2. A larger number of subsets is associated with a larger estimate for the EVPPI and so it is important to assess the correct number of subsets for each parameter individually. 

For the two-step MC procedure the outer loop had 1000 observed posterior samples for all the parameters. As the case study is a Bayesian model, the inner loop of the 2 step MC procedure used MCMC methods to draw a sample from the joint posterior distribution for all the parameters. To ensure convergence and reduce autocorrelation, we took 2 chains of 100,000 samples each for the inner loop, with a burn-in of 9,500 and thinning of 500. Therefore, the eventual sample of the known-distribution net benefits had 1000 observations. The Monte Carlo error on this estimate is still quite large; ideally we would have sampled more observations in the inner loop. This was, however, too large a computational burden. To calculate the value of the current information, we sampled from our MCMC procedure 5,000,000 times, with a burn-in of 9,500 and then thinned to 50,000. This sample size gives a Monte Carlo error on the estimate for the current information around 0.04. As only two treatment options are considered, the incremental net benefit was stored at the end of each loop. The maximum between the stored value and 0 was found for each loop.

Finally, the two different non-parametric regression methods suggested by Strong et al.\ \cite{StrongOakley:2014} were used to calculate the EVPPI by approximating the inner expectation in (\ref{EVPPI}). This analysis was carried out using the functions in \cite{Strong:2012:Code}. For GAM regression, the functions use spline regression as smoothing functions. The \texttt{R} function \verb"gam"  is used by \cite{Strong:2012:Code} to fit the spline regression and \verb"gam"  chooses the dimension of the underlying basis functions. Therefore, no additional input is needed for single parameter GAM regression.  For Gaussian Processes, Strong et al.~use a specific model structure to allow for the use of analytical results, therefore, no model structure needs to be specified by the user and functions are provided to calculate the EVPPI with no additional inputs.

\begin{table}[!h]
\centering
\begin{tabular}{|p{7.35cm}|c|c|c|c|c|c|}
\hline
\multicolumn{2}{|c}{\multirow{2}{*}{Parameter}}&\multicolumn{5}{|c|}{Methods}\\ \cline{3-7}
\multicolumn{2}{|c|}{ } & SO&SAD&GP&GAM&MC\\
\hline
\multirow{7}{*}{Probability of clinical outcomes}&$\beta_1$&1.15&1.17&1.11&1.11&1.06\\
&$\beta_2$&0.26&0.23&0.11&0.11&0.25\\
&$\beta_3$&0.24&0.15&0.03&0.03&0.07\\
&$\beta_4$&0.14&0.13&0.08&0.09&0.05\\
&$\beta_5$&0.28&0.25&0.07&0.12&0.06\\
&$\beta_6$&0.27&0.29&0.18&0.21&0.21\\
&$\beta_7$&0.35&0.4&0.3&0.3&0.25\\
\hline
Number of drugs prescribed&$\delta$&0.3&0.22&0.11&0.03&0.06\\
\hline
Probability of taking time off work&$\eta$&0.11&0&0.04&0&0.06\\
\hline
\multirow{2}{*}{Probability of receiving drugs}&$\gamma_1$&0.19&0.17&0.11&0.11&0.06\\
&$\gamma_2$&0&0&0.01&0.01&0.06\\
\hline
Number of days off work&$\lambda$&0.41&0.44&0.41&0.43&0.43\\
\hline
Reduction in infection rate due to vaccine&$\rho$&0.3&0.3&0.28&0.28&0.28\\
\hline
Probability of taking OTC medication&$\xi$&0.24&0.13&0.03&0.03&0.06\\
\hline
\multirow{5}{*}{QALY cost for each state}&$\omega_1$&0.55&0.57&0.53&0.53&0.63\\
&$\omega_4$&0.2&0.23&0&0&0.06\\
&$\omega_5$&0.11&0.17&0.05&0.05&0.06\\
&$\omega_6$&0.18&0.16&0.12&0.12&0.13\\
&$\omega_7$&0.18&0.29&0.14&0.16&0.15\\
\hline
Vaccine coverage rate&$\phi$&0.13&0&0.02&0.02&0.06\\
\hline
\multirow{8}{*}{Cost of resources}&$\psi_1$&0.17&0.17&0.13&0.13&0.07\\
&$\psi_2$&0.1&0.14&0.02&0.02&0.06\\
&$\psi_3$&0.37&0.37&0.31&0.31&0.29\\
&$\psi_4$&0.46&0.52&0.35&0.35&0.42\\
&$\psi_5$&0.44&0.38&0.24&0.38&0.25\\
&$\psi_6$&0.13&0.15&0.05&0.05&0.07\\
&$\psi_7$&0.2&0&0&0&0.07\\
&$\psi_8$&0.08&0&0&0&0.07\\
\hline
\end{tabular}
\caption{The EVPPI values for the parameters in the Vaccine example using 5 different methods discussed in this paper - Strong and Oakley single parameter estimation (SO), Sadatsafavi \textit{et al.} single parameter estimation (SAL), Gaussian Process (GP) and GAM regression (GAM) and two-step Monte Carlo simulation (MC)}
\label{EVPPI-table}
\end{table}
\subsection{Results}
Table \ref{EVPPI-table} displays the single parameter EVPPI estimates for the five different methods. All the methods give approximately the same results for each parameter. The two parameters with the highest EVPPI are the same across all five methods: $\beta_1$ and $\omega_1$. There is then some discrepancy in the ordering, but the results only differ slightly, certainly in comparison to the Monte Carlo error of the current information which is 0.04.

\begin{figure}
\centering
\includegraphics[width=10cm]{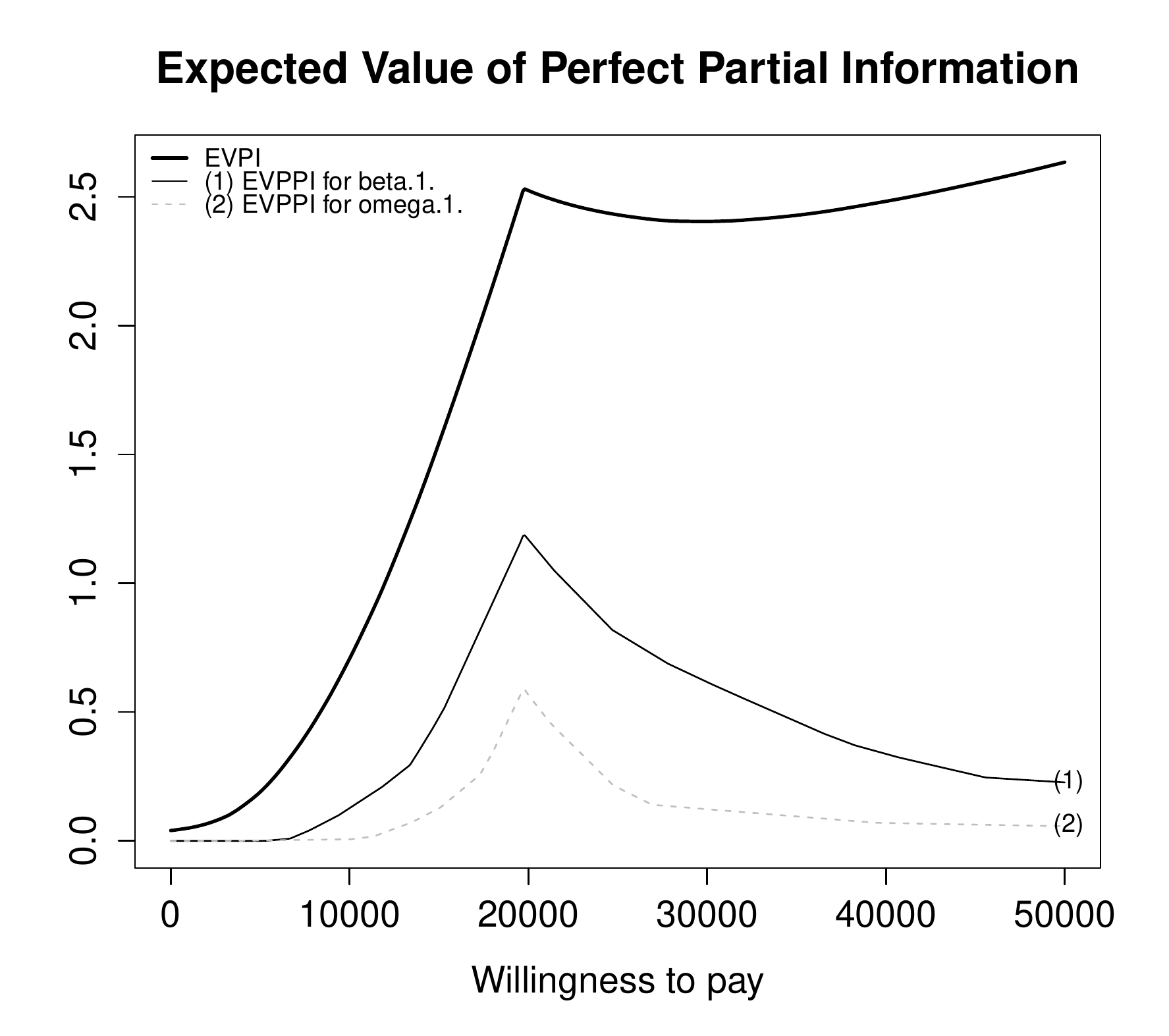}
\caption{The EVPI and EVVPI for $\beta_1$ and $\omega_1$ for willingness-to-pay thresholds varying between 0 and 50000. The Sadatsafavi et al.~method with one decision change is used to calculate the single-parameter~EVPPI.}
\label{graph}
\end{figure}

Figure \ref{graph} shows the EVPI and the EVPPI for $\beta_1$ and $\omega_1$ for different values of the willingness-to-pay, $k$. The analysis in this paper was undertaken for $k=20000$ which is near the break-even point between the two treatment options. Both the EVPI and the EVPPI reach a peak when the optimal decision under current uncertainty changes. This is because the uncertainty is greatest at the point when both strategies are equal. As the EVPPI is very small for most of the parameters, even for a willingness-to-pay close to the break-even point, it is clear that uncertainty has little impact on the decision in this example. If the values of $k$ were to change greatly in this example, we would be approximating very small EVPPI values and the MC error for the 2 step MC method would completely dominate the estimates.

These results show that the approximation methods offer similar results for single parameter EVPPI estimates and can therefore all be used to estimate the EVPPI.  The variation between the methods is likely to be due to the inherent nature of approximation. It is important to note that the accuracy of the estimate of the EVPPI should only be reported to 1 or 2 decimal places even when the analysis gives a much greater accuracy of calculation as all these methods only approximate the true EVPPI.
\begin{figure}[!h]
\centering
\includegraphics[width=13.5cm]{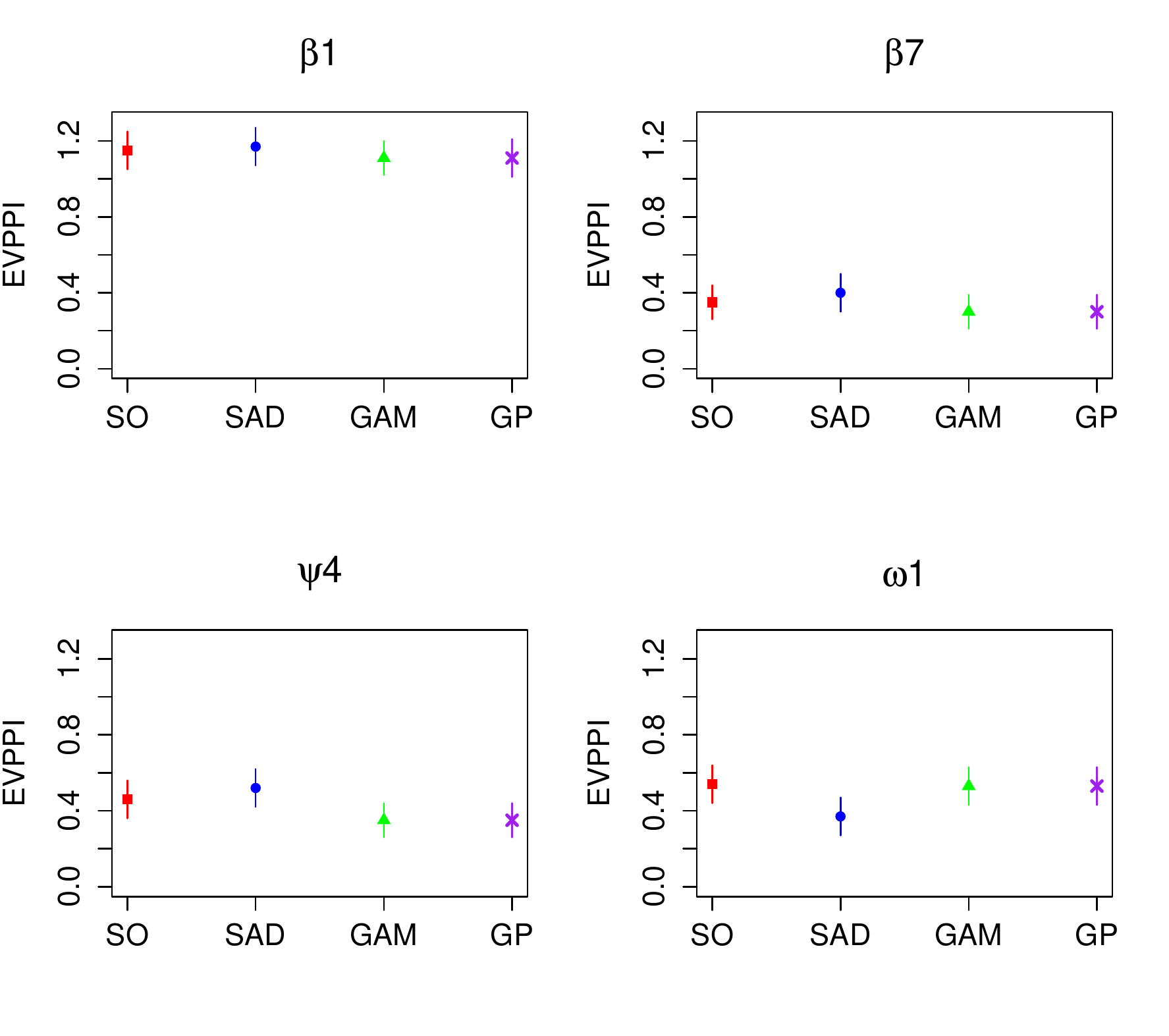}
\caption{The mean EVPPI estimate and its standard error for four estimation techniques - Strong and Oakley (SO), Sadatsafavi \textit{et al.} (SAD), GAM and Gaussian Process (GP) regression -- for the four most important parameters}
\label{SE}
\end{figure}

Given that all five methods considered produce broadly similar estimates for the single parameter EVPPI, it seems pertinent to investigate the standard error of the EVPPI calculated using these methods, as well as the computation time required to produce the estimate. The single-parameter estimation methods and the GAM regression take negligible time to find single parameter EVPPI estimates. The GP regression is also fast for single parameter EVPPI, taking around 2-4 seconds. The Monte Carlo procedure, however, takes around 55 minutes to calculate a single parameter EVPPI estimate. Therefore, all the approximation methods offer significant computational savings.

Taking these computational times into account, the four estimation methods are approximately equivalent, although the GP does take noticeably longer than the other methods. Therefore, an investigation of the standard error was undertaken to see if one method outperforms the others. Figure~\ref{SE} shows the standard error for all the methods (excluding the Monte-Carlo procedure) for the four parameters with the highest EVPPI. Clearly, the standard error is similar across all four methods and therefore there is little difference between the estimation properties of all 4 methods.  

\section{Discussion}\label{discussion}

This paper presents the current methods developed for approximating the EVPPI. It particularly focuses on four EVPPI estimation methods that have been developed since an earlier review paper on methods for estimating EVPPI \cite{CoyleOakley:2008}. We compared the estimation properties of these 4 estimation methods taking into the account the computational time required to approximate the EVVPI and the standard error of the estimate. Despite varying in complexity, these 4 new methods gave consistent results when compared with the 2-step MC procedure that derives directly from the definition of the EVPPI and can thus be considered to provide the ``true'' value. The four methods also have a similar standard error and computational time. Thus, their performance should be compared on the basis of simplicity of implementation.

Two of the methods can only be used for single parameter EVPPI which clearly limits their applicability. Additionally, a thorough understanding of the methods and the respective tools for identifying the correct input values (visual tool, bias reduction) is essential before using of these methods. The \verb"BCEA" \cite{BCEA:2013} package in \texttt{R} allows users to calculate single parameter EVPPI using either method. However, the input values are still needed and a naive implementation of this function will lead the experimenter to grossly over/under-estimate the single parameter EVPPI. Therefore, these methods are largely superseded by the non-parametric regression method which are simpler to use as well as being applicable to multi-parameter~EVPPI. 

%%%%These methods vary in complexity and computational time but gave consistent results. The methods should, therefore, be compared on their computational time and complexity of implementation. The two-step Monte Carlo method is impractical, even in this relatively simple example the computational time for a single parameter EVPPI estimate is around 1 hour. Additionally, the MC error on this estimate is high compared to the value of the underlying EVPPI. Therefore, as noted in previous papers, \cite{Sadatsafavietal:2013,StrongOakley:2014} approximation methods should be used. 

%%All four approximation methods have a similar standard error. Therefore, there is little difference between the methods in terms of accuracy. Both the methods for single parameter estimation produce similar results and can thus be used quickly and efficiently when we only wish to calculate single parameter EVPPI. However, both these methods are sensitive to changes in the input values, $M$.  These methods should therefore be implemented with care and shouldn't be used without some discussion of these input values. %For example, in order to use the Sadatsafavi \textit{et al.} method we must determine the number of decision changes. If the true number of decision changes is 0 then rightly the EVPPI should also be 0. If however, we mistakenly believe there is one decision change then we will observe a positive EVPPI. 

As the GAM regression requires negligible computational effort for single-parameter EVPPI, it is the best method for approximating the EVPPI under these circumstances. However, for larger subsets, the interaction structure of the parameter must be specified by the user. Thus, the experimenter needs some understanding of GAM regression. Additionally, GAM regression is mathematically unstable for larger subsets of parameters ($\geq$ 6).

Therefore, the slower Gaussian Process regression is currently the only feasible option for larger parameter subsets, especially where  the experimenter has no information about the interaction structures present in the parameter space. Unfortunately the Gaussian Process method takes more computational time as the number of parameters increases and there is a trade-off between the computational time spent fitting the model and the accuracy of the EVPPI estimate. For example, for a PSA data set of size $N$ the computational effort for the Gaussian Process regression method is proportional to $N^3$. Clearly, therefore, the smaller the PSA data set the faster the EVPPI calculation. However, the more data available to fit the non-parametric regression the more accurate the EVPPI estimate. %The function \cite{Strong:2012:Code} that fits Gaussian Process regression calculates an estimate of the standard error of the estimator. One can therefore judge the accuracy of the estimator. 
%%
%%In conclusion, the most promising methods for approximating the EVPPI is the use of non-parametric regression. However, for larger parameter subsets the computational burden can still be substantial especially when the fast GAM regression cannot be used. Therefore, additional research should target reducing this computational time so that fast EVPPI approximation methods are available for all possible parameter subsets.
An extension to the standard GP regression approach exploiting results from the geostatistics literature is being investigated \cite{Heathetal:2015} and results appear to be promising both in terms of accuracy and reduced computational time.

All the methods investigated in this paper can be implemented in suitable software, such as \texttt{R} using the package \verb"BCEA" \cite{BCEA:2013} or using the standalone functions available from the original authors. Therefore, to some extent they can be applied in a ``black-box'' sense; which may lead to a more widespread use of EVPPI in health economic evaluations and allow decision makers to use Value of Information methods to inform their decisions. 

\bibliographystyle{plain}
\bibliography{bib}

\appendix
\section{Visual Tool for Sadatsafavi et al. method}\label{tool}

The Sadatsafavi \textit{et al.} \cite{Sadatsafavietal:2013} method for single parameter EVPPI estimation relies on an input value determining the number of changes in the optimal decision caused by the parameter of interest. To aid the choice of this value Sadatsafavi \textit{et al.} suggest a visual tool that, for every pair of decisions $t, t'$, can be used to determine the number of changes in optimal decision. Firstly we must order our PSA data in the parameter of interest, $\phi$. Then, as a function of $\phi$ we define \begin{equation}\hat{C}_{t,t'}(\phi) = \frac{1}{S} \sum_{s=1}^S I(\phi^s <\phi) (\mbox{NB}_t(\phi^{s}) - \mbox{NB}_{t'}(\phi^{s}))\end{equation} where $\mbox{NB}_t(\phi^{s})$ is the observed utility value for the $s$-th lowest value of $\phi$ for treatment $t$. The extrema in this estimator will correspond to a change in optimal decision. For example, if $t$ is the optimal decision until $\phi=1$ then as $\phi$ increases from 0 to 1, only positive terms will be summed, then as the optimal decision changes after 1, negative terms will be summed and therefore the value of $\hat C$ will decrease. 

\begin{figure}[!h]
\centering
\includegraphics[width=7.25cm]{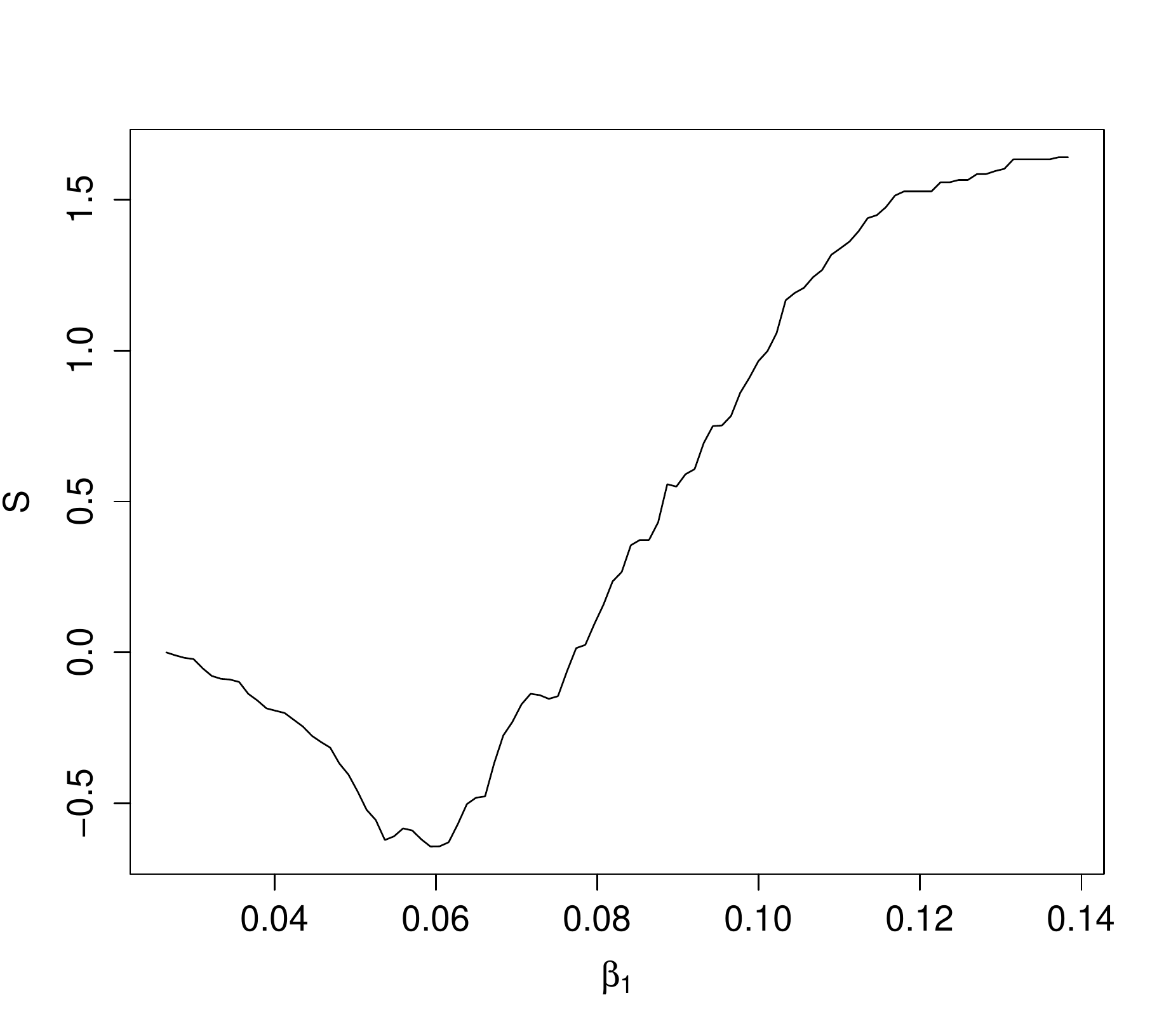}
\includegraphics[width=7.25cm]{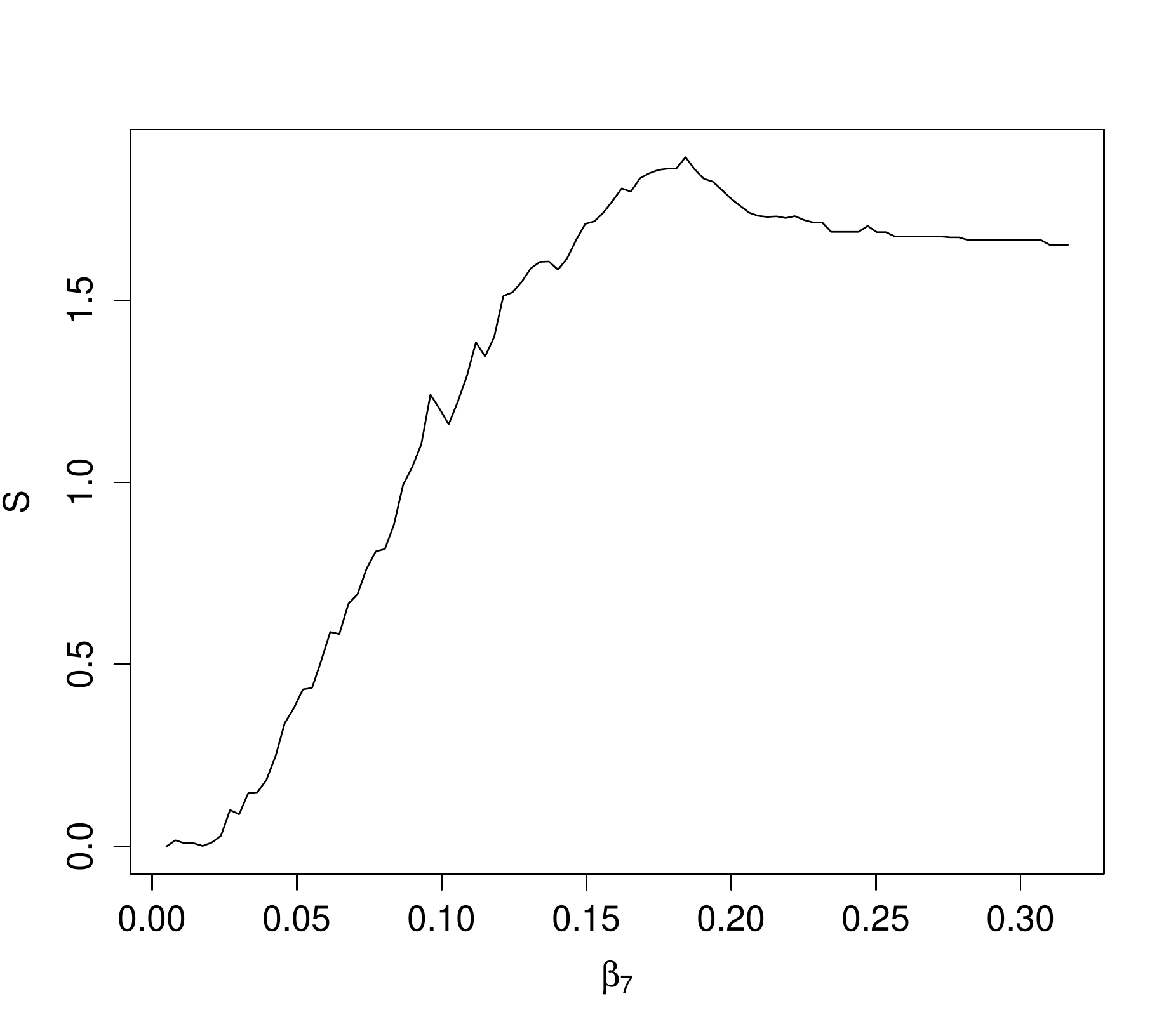}
\includegraphics[width=7.25cm]{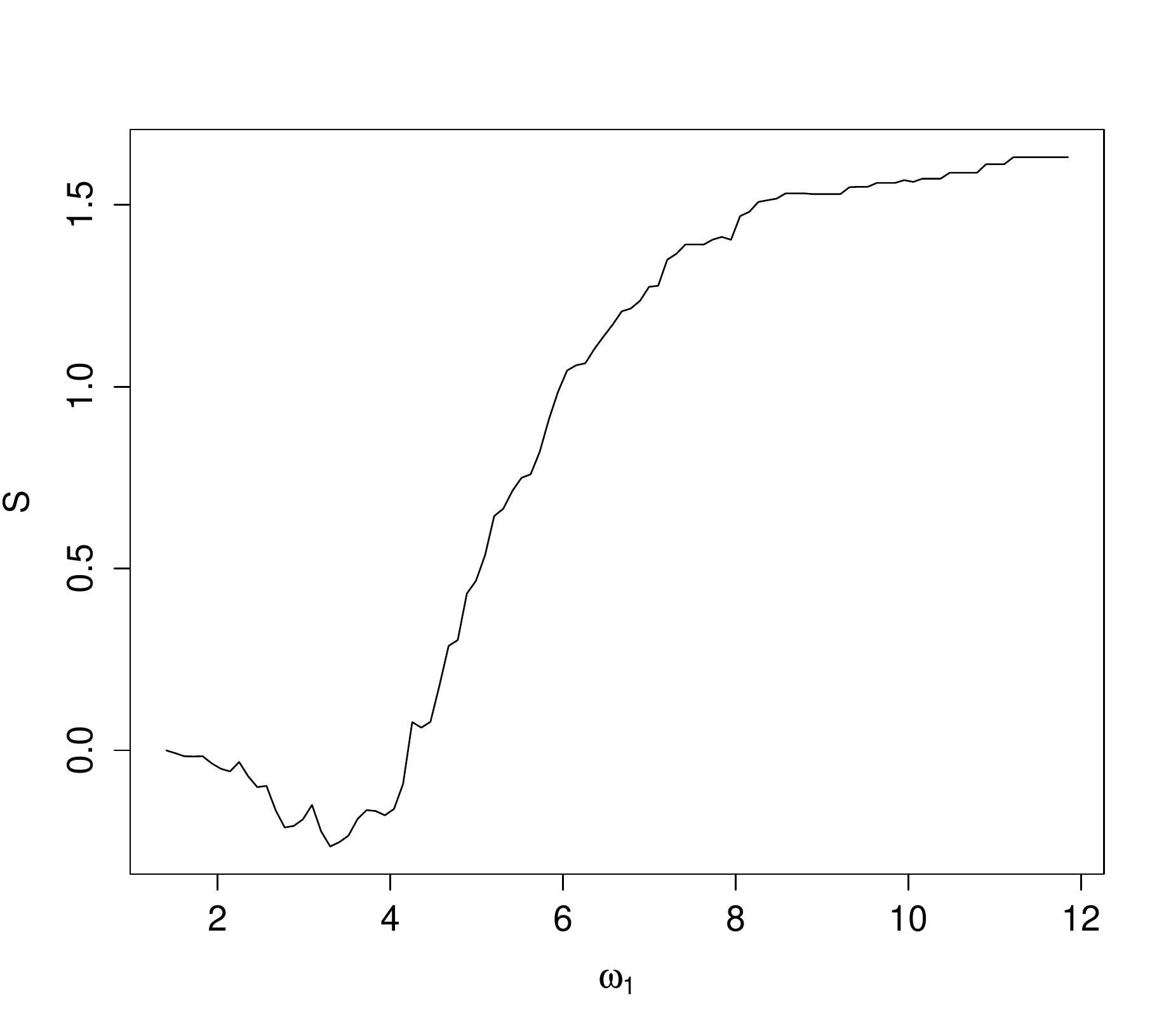}
\includegraphics[width=7.25cm]{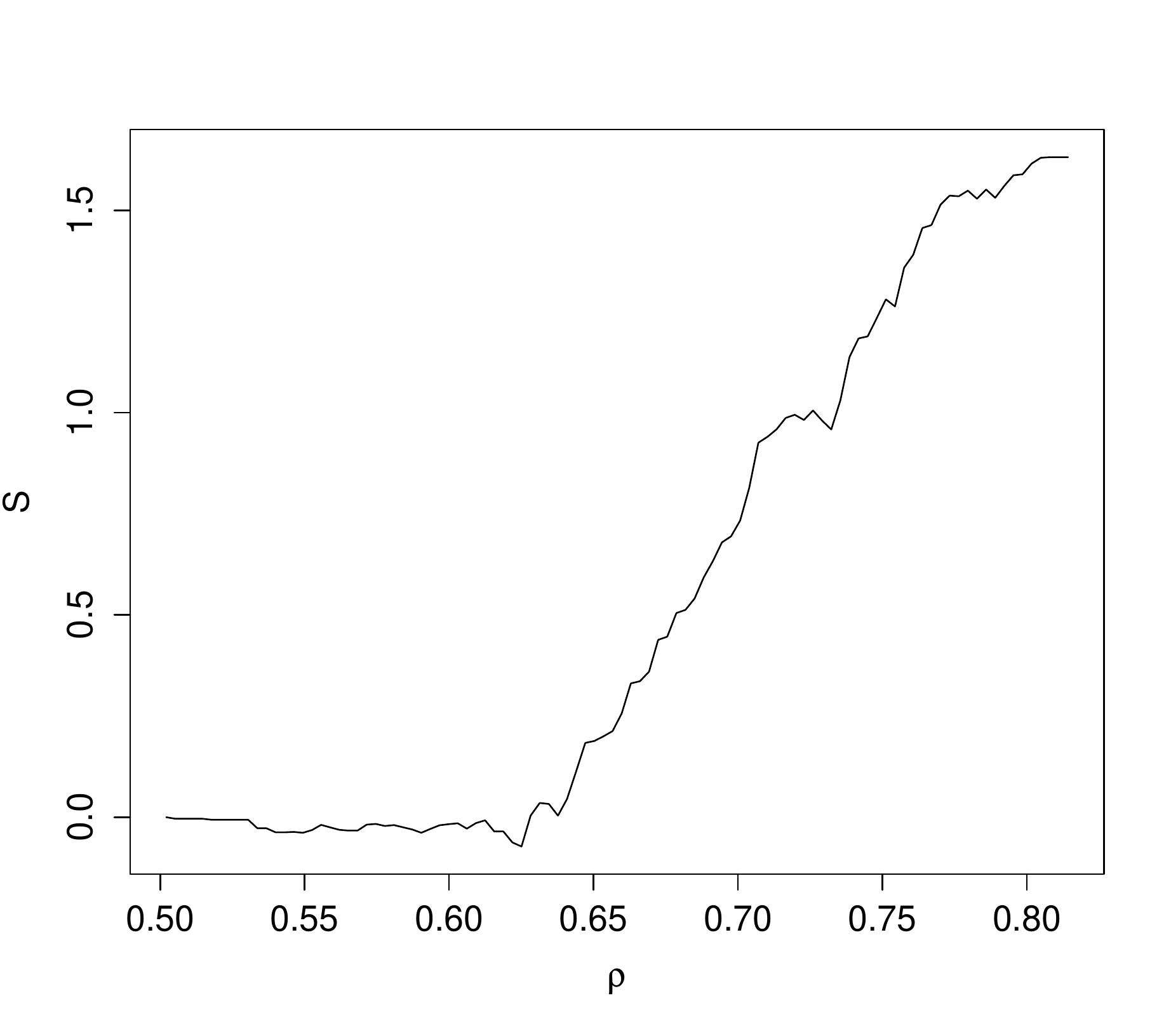}
\label{visual}
\caption{The visual tool for determining the number of time the optimal decision changes for four different parameters: $\beta_1, \beta_7, \omega_1$ and $\rho$.}
\end{figure}

We plot the observed values for $\hat C$ against the parameter of interest. This allows us to visually search for maxima or minima. As we only have empirical values and not a fully-specified function, a computer cannot find the extrema. Using the Vaccine case study, Figure \ref{visual} shows the visual tool for four the parameters to illustrate the difficulty faced in deciding whether a parameter affects the optimal decision. 

Firstly, we consider the tool for $\beta_1$ (the top left-hand corner). Clearly, this plot has one extremum, around 0.06. There is a very obvious change in direction for the graph. This is the same for the graph for $\omega_1$, where the direction change is clearly around 3.5. These two parameters are those with the highest EVPPI across all four methods, so the visual tool is highly effective for parameters with high EVPPI.

Determining extrema is more challenging for $\beta_7$ and $\rho$. The plot of $\beta_7$ rises steadily before possibly reaching a peak and decreasing slightly around 0.175. There may, therefore, be a decision change but is less obvious than for $\beta_1$ and $\omega_1$. For $\rho$ it is challenging to ascertain whether we have an extremum at 0.625. There does seem to be some slight decrease to 0.0625 but it is very unclear. For this analysis, it was decided that $\rho$ does cause the optimal decision to change giving an EVPPI of around 0.3. 
\end{document}